\documentclass[twocolumn]{jpsj2} 
\title{Replica Analysis of a Preferential Urn Model}

\author{Jun \textsc{Ohkubo}$^{1,2}$\thanks{E-mail address: jun@smapip.is.tohoku.ac.jp}, 
Muneki \textsc{Yasuda}$^{1}$
and Kazuyuki \textsc{Tanaka}$^{3}$
}

\inst{$^{1}$Department of System Information Sciences, Graduate School of Information Sciences,\\
Tohoku University, 6-3-09, Aramaki-Aza-Aoba, Aoba-ku, Sendai 980-8579, Japan\\
$^{2}$JSPS Research Fellow\\
$^{3}$Department of Applied Information Sciences, Graduate School of Information Sciences,\\
Tohoku University, 6-3-09, Aramaki-Aza-Aoba, Aoba-ku, Sendai 980-8579, Japan\\
}

\abst{
We analyze a preferential urn model with randomness using the replica method.
The preferential urn model is a stochastic model based on the concept ``the rich get richer.''
The replica analysis clarifies that
the preferential urn model with randomness shows a fat-tailed occupation distribution.
We discuss the intuitive picture of the urn model, a relationship between quenched and annealing systems, 
and a trial for the application of the preferential urn model to econophysics.
The analytical treatments and results would be useful for various research fields
such as complex networks, stochastic models, and econophysics.
}

\kword{urn model, replica analysis, stochastic process, statistical mechanics, Ehrenfest class, 
Pareto's distribution, econophysics}

\begin{document}
\maketitle

\section{Introduction}

Urn models have been used in order to explain various physical phenomena such as quantum gravity and glassy dynamics
\cite{Ritort1995,Bialas1997,Drouffe1998,Bialas1998,Bialas2000,Godreche2001,Leuzzi2002,Burda2002}.
Because of their tractability for analytical calculations,
the urn models have been studied extensively in statistical physics and probability theory.
The examples of the urn models are the backgammon model \cite{Ritort1995}
and the zeta urn model \cite{Bialas1997};
these models can be treated analytically.
Furthermore, it has been shown that these urn models relate to the zero-range process
which is also famous in the statistical physics; see a review by Evans and Hanney \cite{Evans2005}.

The preferential urn model\cite{Ohkubo2005c} is also one of the examples of the urn models.
It has been revealed that the preferential urn model shows a fat-tailed occupation distribution
only in the case with randomness; the meanings of the ``randomness'' will be explained later in the present paper.
The fat-tailed behavior is similar to the Parato's distribution \cite{Newman2005}.
The preferential urn model stems from research on complex networks \cite{Ohkubo2005c},
and because of the simplicity,
the model would have a wide range of applications in research on not only complex networks,
but also many other fields such as econophysics and social sciences.
In econophysics or social sciences, 
it could be reasonable to consider heterogeneous cases, e.g., 
in which each agent in econophysical systems has a different ability.

In the present paper, an analysis of an urn model with randomness is presented.
While equilibrium properties of urn models without randomness are easily obtained \cite{Godreche2001},
the analysis for the urn models with randomness are not straightforward.
We analyze the preferential urn model 
with randomness by using the replica method \cite{Nishimori2001}.
Furthermore, we will give some discussions for the analytical treatment
and an example of the applications of the preferential urn model for econophysics,
though the example is just a trial one.

The outline of the present paper is as follows.
Section~2 gives a brief introduction to urn models.
In \S~3, we introduce the preferential urn model.
The general discussion for calculating occupation distributions is given in \S~4,
and \S~5 gives the replica analysis for the preferential urn model.
Some discussions for the replica analysis and an example of the applications of the model are given in \S~6.
Finally, we draw the main conclusions in \S~7. 

\section{Backgrounds of Urn Models}

One of the most important properties of the urn models is that the partition function is given by a factorized form
\cite{Godreche2001}.
For example, we consider a system with $N$ urns and $M$ balls which are distributed among $N$ urns.
When we define an energy of each urn as $E(n_i)$
by using the number of balls in the urn, $n_i$,
the total energy of the whole system is given by 
\begin{align}
E(n_1, \dots, n_N) = \sum_{i=1}^N E(n_i).
\end{align}
The partition function of the system is calculated from
\begin{align}
Z = \sum_{n_1=1}^\infty \cdots \sum_{n_N=1}^\infty p_{n_1} \cdots p_{n_N} \delta\left( \sum_{i=1}^N n_i , M \right),
\label{intro_partition_function_monkey}
\end{align}
where 
\begin{align}
p_{n_i} = e^{-\beta E(n_i)}
\label{intro_boltzmann}
\end{align}
is the unnormalized Boltzmann weight attached to urn $i$, and $\beta$ is an inverse temperature.
The Kronecker delta of eq.~\eqref{intro_partition_function_monkey} 
stems from the restriction of the total number of balls.
We can obtain various physical quantities 
by using the partition function of eq.~\eqref{intro_partition_function_monkey};
for example, we can calculate an equilibrium occupation probability
$P(k)$ with which an urn has $k$ balls in the equilibrium state.

The above example, i.e., eq.~\eqref{intro_partition_function_monkey}, 
corresponds to \textit{monkey class} of urn models \cite{Godreche2001}.
In the Monkey class, an urn is selected at random, from which any ball is drawn,
and transfer the drawn ball to the other urn;
this dynamics corresponds to the image of a monkey playing at exchanging balls among urns.
In other words, each ball is indistinguishable in the monkey class,
so that the equilibrium statistics of the monkey class is considered as Bose-Einstein statistics (quantum).
On the other hand, there is the other type of urn models, so-called \textit{Ehrenfest class},
which corresponds to the Maxwell-Boltzmann statistics (classical).
In the Ehrenfest class, balls within an urn are distinguishable,
so that the partition function of the Ehrenfest class is written as
\begin{align}
Z = \sum_{n_1=1}^\infty \cdots \sum_{n_N=1}^\infty 
\frac{p_{n_1}}{n_1!} \cdots \frac{p_{n_N}}{n_N!} \delta\left( \sum_{i=1}^N n_i , M \right),
\label{intro_partition_function_Ehrenfest}
\end{align}
which involves inverse factorials $1/(n_i!)$; 
it is different from the partition function of the
monkey class of eq.~\eqref{intro_partition_function_monkey}.

The choice of the energy of each urn is arbitrary,
and hence one can consider a lot of models
so that a selected model is suitable for his/her own problem.

\section{Preferential Urn Model}

\begin{figure}
\begin{center}
  \includegraphics[width=7.5cm,keepaspectratio,clip]{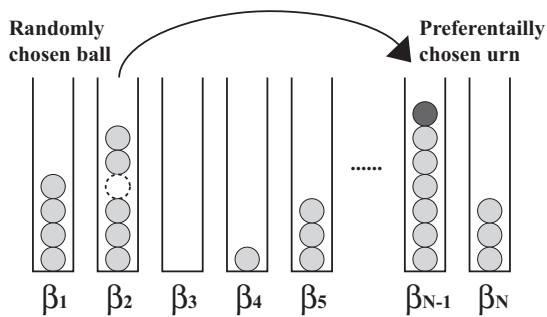} 
\caption{
Illustration of the preferential urn model.
A ball is drawn at random,
and then the drawn ball is moved to an urn chosen preferentially.
}
\label{fig_model}
\end{center}
\end{figure}

We consider a system of $M$ balls distributed among $N$ urns,
and denote the density of the system as $\rho = M/N$.
The number of balls contained in urn $i$ is denoted by $n_i$, and hence $\sum_{i=1}^N n_i = M$.
For the preferential urn model, we define the following energy of each urn:
\begin{align}
E(n_i) = - \ln (n_i!).
\label{eq_energy}
\end{align}
From the definition of the energy,
it is clear that each urn tends to get more and more balls
because an urn is stable when the urn obtains a large number of balls.
Note that there is a restriction for the total number of balls,
so that an urn cannot get an infinitely large number of balls.
It would be clarified that the transition rate in the urn model with the above definition of the energy obeys
a preferential concept, as described later,
and hence the urn model is called the \textit{preferential urn model}.
Furthermore, we here assume that
the preferential urn model belongs to the Ehrenfest class.

The unnormalized Boltzmann weight of eq.~\eqref{intro_boltzmann} is written by
\begin{align}
p_{n_i} = \left( n_i! \right)^{\beta_i},
\end{align}
where $\beta_i$ is an inverse temperature of urn $i$.
Note that each urn has own inverse temperature and
each inverse temperature could be different from each other;
we denote the distribution of the inverse temperatures as $\phi(\beta)$.
Though it might seem strange in terms of physics that each urn has a different temperature,
there should be no restriction for a homogeneity of temperature 
in information physics, social sciences, and econophysics
(Or one could say that the system is in nonequilibrium.)
We call the inverse temperature ``the inverse \textit{local} temperature'' because of its locality.
It will be shown in \S~4 and \S~5 that 
the locality is important for generating a fat-tailed occupation distribution,
and, furthermore, 
these inverse local temperatures allow us to understand results of an equilibrium occupation distribution intuitively.

Using the heat-bath rule,
we derive the transition rate $W_{n_i \to n_i + 1}$ from the state $n_i$ to $n_i + 1$ as follows \cite{Drouffe1998}:
\begin{align}
W_{n_i \to n_i + 1} & \propto  \frac{p_{n_i+1}}{p_{n_i}} \nonumber \\
&= (n_i+1)^{\beta_i}.
\end{align}
The transition rate indicates that an urn with more balls tends to get more balls,
and this means the preferential concept.
The dynamics of the preferential urn model is illustrated in Fig.~1.

In summary, the construction of the preferential urn model is as follows:
\begin{enumerate}
\item[(i)]
Set $N$ urns and $M$ balls.
The $M$ balls are distributed among the $N$ urns,
and the initial distribution is arbitrary (usually distributed at random).
Each inverse local temperature $\beta_i$ is chosen by using a distribution $\phi(\beta)$.
\item[(ii)]
Choose a ball at random (the Ehrenfest class).
\item[(iii)]
Move the drawn ball to an urn selected by using the transition rate $W_{n_i \to n_i + 1}$.
\item[(iv)]
Repeat the procedures (ii) and (iii) until the system reaches an equilibrium state.
\end{enumerate}

If all urns have the same temperatures, i.e., $\beta_i = \beta$ for all $i$,
equilibrium occupation distributions are easily calculated \cite{Bialas1997,Godreche2001,Ohkubo2005c}.
For instances, the equilibrium occupation distribution in the high-temperature limit ($\beta_i \to 0$ for all $i$)
is a Poisson distribution \cite{Godreche2001,Ohkubo2005c}.
However, it is generally difficult to calculate equilibrium occupation distributions of urn models with randomness. 
In the following sections,
we give an analytical treatment to calculate the equilibrium occupation distribution of the preferential urn model
with randomness.

\section{Occupation Distribution}

In this section, we give a general discussion for calculating occupation distributions,
and it will be revealed that the case with randomness is difficult to be analyzed.
In the next section, the replica analysis is given,
and an explicit form of the occupation distribution will be obtained in the case with randomness.

By using the integral representation of the Kronecker delta, $2 \pi i \delta(a,b) = \oint \textrm{d}z \, z^{a-b-1}$,
the partition function of eq.~\eqref{intro_partition_function_Ehrenfest} is calculated as follows:
\begin{align}
Z_1 &= \sum_{n_1 = 0}^\infty \dots \sum_{n_N = 0}^\infty (n_1!)^{\beta_1 - 1}
 \dots (n_N!)^{\beta_N - 1} \nonumber \\
& \qquad \times \frac{1}{2\pi i}\oint\!\textrm{d}z \, z^{\sum_{i=1}^N n_i - M - 1}.
\label{eq_Z_N_M}
\end{align}
Note that all values of the inverse local temperatures
$\{\beta_1, \dots, \beta_N\}$ are clamped at certain values in eq.~\eqref{eq_Z_N_M}.
Here, we define $f_{k}^{(\beta_1)}$ as an equilibrium occupation probability of urn $1$ 
with the inverse local temperature $\beta_1$ in a certain configuration:
\begin{align}
f_{k}^{(\beta_1)} &= \frac{1}{Z_1}  
\sum_{n_1 = 0}^\infty \dots \sum_{n_N=0}^\infty \delta(n_1,k) \frac{p_{n_1}}{n_1!} \dots \frac{p_{n_N}}{n_N!} 
\nonumber \\
& \qquad \times \delta\left( \sum_{i=1}^{N} n_i , M\right).
\label{eq_fk_not_simple}
\end{align}
Then, $f_{k}^{(\beta_1)}$ indicates the probability with which urn $1$ has $k$ balls in a certain configuration.
Next, we define $Z_2$ as
\begin{align}
Z_2 &= 
\sum_{n_2 = 0}^\infty \dots \sum_{n_N = 0}^\infty
 (n_2!)^{\beta_2 - 1} \dots  (n_N!)^{\beta_N - 1} \nonumber \\
& \qquad \times \oint \frac{\textrm{d}z}{2\pi i}
z^{k + \sum_{i=2}^N n_i - M - 1} ,
\end{align}
and simplify $f_{k}^{(\beta_1)}$ of eq.~\eqref{eq_fk_not_simple} as follows:
\begin{align}
f_{k}^{(\beta_1)} = (k!)^{\beta_1 - 1} \frac{Z_2}{Z_1}.
\label{eq_urn1_clamped}
\end{align}

In order to obtain the equilibrium occupation distribution of urn $1$ with $\beta_1$ 
in the configurational average, $P(k, \beta_1)$,
we average $f_{k}^{(\beta_1)}$ over the configuration $\{\beta_2, \cdots, \beta_N \}$:
\begin{align}
P(k, \beta_1)
&= \left\langle f_{k}^{(\beta_1)} 
 \right\rangle_{\{\beta_2, \dots, \beta_N\}} \nonumber \\
&= (k!)^{\beta_1 - 1} 
 \left\langle \frac{Z_2}{Z_1}  \right\rangle_{\{ \beta_2, \dots, \beta_N \}},
\label{eq_p_k_0}
\end{align}
where $\langle \cdots \rangle_{\{ \beta_2, \dots, \beta_N \}}$
means that the configurational average is taken only over $\{\beta_2, \dots, \beta_N\}$;
the configurational average of a quantity $A(\beta_1, \beta_2, \cdots, \beta_N)$ is defined as
\begin{align}
&\left\langle A(\beta_1, \beta_2, \dots, \beta_N) \right\rangle_{\{\beta_2, \dots, \beta_N\}}
\nonumber \\
&\quad = \int \phi(\beta_2) \textrm{d}\beta_2 \cdots \int \phi(\beta_N) \textrm{d}\beta_N \,
A(\beta_1, \beta_2, \dots, \beta_N).
\end{align}
Finally, we take the configurational average of $P(k,\beta)$ for $\beta$.
Then, the equilibrium occupation distribution is calculated from
\begin{align}
P(k) &= \int \textrm{d}\beta \,\, \phi(\beta) P(k,\beta).
\label{eq_distribution_1}
\end{align}

In eq.~\eqref{eq_p_k_0},
it is difficult to take the configurational average of the numerator and denominator at the same time.
Therefore, we use replica analysis for evaluate the equilibrium occupation distribution.

\section{Replica Analysis}
The replica method \cite{Nishimori2001} is one of the powerful methods for analysing random systems.
In the replica analysis, we prepare $m$ replicas of the original system,
and evaluate the configurational average of the partition functions, assuming the thermodynamic limit.
After that, the limit $m \to 0$ is taken,
which has successfully clarified typical properties of many random systems.
While mathematical validation of the replica analysis still remains unresolved,
the replica analysis has been empirically recognized as a reliable scheme in physics;
for instance, we can use the replica analysis in order to calculate a free energy or order parameters.
In this section,
we use the replica analysis to obtain the equilibrium occupation distribution of the preferential urn model
with randomness.
The aim of the present paper is to propose the analytical treatment for the preferential urn model with randomness,
and hence we here restrict ourselves only to the case 
in which the distribution of the inverse local temperatures, $\phi(\beta)$,
follows an uniform distribution, i.e., $\phi(\beta) = 1$ for $\beta \in [0,1]$.

First, we assume the following expression in order to calculate eq.~\eqref{eq_p_k_0}:
\begin{align}
\left\langle f_{k}^{(\beta_1)} 
\right\rangle_{\{\beta_2, \dots, \beta_N\}} 
= \exp\left(  
\left\langle \ln f_{k}^{(\beta_1)} 
\right\rangle_{\{\beta_2, \dots, \beta_N\}} 
\right).
\label{eq_p_k}
\end{align}
The expression is obtained from the \textit{self-averaging property}\cite{Nishimori2001}
because of the equilibrium occupation probability $f_{k}^{(\beta_1)}$ is intensive.
Therefore, it is necessary only to calculate the configurational average of 
$\ln f_{k}^{(\beta_1)}$; this term is calculated from
\begin{align}
 \left\langle \ln f_{k}^{(\beta_1)} \right\rangle_{\{ \beta_2, \dots, \beta_N\}}
= &(\beta_1 - 1) \ln k! 
- \left\langle \ln Z_1 \right\rangle_{\{ \beta_2, \dots, \beta_N\}} \nonumber \\
&+ \left\langle \ln Z_2 \right\rangle_{\{ \beta_2, \dots, \beta_N\}}. 
\label{eq_ln_f}
\end{align}
In the replica analysis,
we use the following manipulations to evaluate the configurational average:
\begin{align}
\left\langle \ln Z_i \right\rangle_{\{ \beta_2, \dots, \beta_N\}}
=& \lim_{m\to 0} \left( \frac{\left\langle Z_i^m \right\rangle_{\{ \beta_2, \dots, \beta_N\}} - 1}{m}
\right) \nonumber \\
&\qquad \qquad \qquad \qquad (i = 1,2).
\label{eq_replica}
\end{align}
We should calculate the following quantities to evaluate eq.~\eqref{eq_ln_f} 
using eq.~\eqref{eq_replica}.
Therefore, we first calculate the following quantities:
\begin{align}
&\left\langle Z_1^m \right\rangle_{\{ \beta_2, \dots, \beta_N\}} \nonumber \\
&\quad = \oint \prod_{\alpha = 1}^m  \left\{ \frac{\textrm{d}z_\alpha}{2\pi i}  z_\alpha^{-(M+1)}
G(z_\alpha)^{N-1} H(z_\alpha, \beta_1) \right\},
\label{eq_replica_1} \\
&\left\langle Z_2^m \right\rangle_{\{ \beta_2, \dots, \beta_N\}} \nonumber \\
&\quad = \oint \prod_{\alpha = 1}^m  \left\{ \frac{\textrm{d}z_\alpha}{2\pi i}  z_\alpha^{-(M+1-k)}
G(z_\alpha)^{N-1} \right\},
\label{eq_replica_2}
\end{align}
where 
\begin{align}
G(z) &= \int_0^1 \textrm{d}\beta  \,\, \phi(\beta) \sum_{n=0}^\infty (n!)^{\beta-1} z^n  \nonumber \\
&= \int_0^1 \textrm{d}\beta \,\, \sum_{n=0}^\infty (n!)^{\beta-1} z^n 
\label{eq_g0}
\end{align}
and
\begin{align}
H(z,\beta) = \sum_{n=0}^\infty (n!)^{\beta-1} z^{n}.
\end{align}
As mentioned above, we here treat the case with uniform randomness, i.e., $\phi(\beta) = 1, \,\, \beta \in [0,1]$,
so that the second line of eq.~\eqref{eq_g0} can be derived.
Next, we assume \textit{replica symmetry},
i.e., $z_\alpha = z$,
and derive the replica symmetric solution.
Then, by using the saddle-point method, eqs~\eqref{eq_replica_1} and \eqref{eq_replica_2} become as follows:
\begin{align}
&\left\langle Z_1^m \right\rangle_{\{ \beta_2, \dots, \beta_N\}} \nonumber \\
&\quad \simeq  \exp\left\{ 
-m \ln z_\textrm{s} - m\ln G(z_\textrm{s}) + m \ln H(z_\textrm{s},\beta_1) \right. \nonumber \\
& \quad \qquad \left. - m\rho N \ln z_\textrm{s} + mN \ln G(z_\textrm{s}) + m\ln C
\right\} \nonumber \\
&\quad \simeq  1 - m\ln z_\textrm{s} - m\ln G(z_\textrm{s}) + m \ln H(z_\textrm{s},\beta_1) \nonumber \\
& \quad \qquad - m\rho N \ln z_\textrm{s} + mN \ln G(z_\textrm{s}) + m\ln C,
\label{Z_1_approximate}
\end{align}
where $C$ is a constant;
\begin{align}
&\left\langle Z_2^m \right\rangle_{\{ \beta_2, \dots, \beta_N\}} \nonumber \\
&\quad \simeq 1 - m(1-k)\ln z_\textrm{s} - m\ln G(z_\textrm{s}) \nonumber \\
& \qquad - m \rho N \ln z_\textrm{s} + mN \ln G(z_\textrm{s}) + m\ln C.
\label{Z_2_approximate}
\end{align}
In deriving the last expression of eqs.~\eqref{Z_1_approximate} and \eqref{Z_2_approximate},
we consider a second-order Taylor expansion because the limit $m \to 0$ will be taken.
The saddle-points $z_\textrm{s}$ in $\left\langle Z_1^m \right\rangle_{\{ \beta_2, \dots, \beta_N\}}$ 
and $\left\langle Z_2^m \right\rangle_{\{ \beta_2, \dots, \beta_N\}}$
are the same and calculated by using the following saddle-point equation:
\begin{align}
\rho &= \frac{z_\textrm{s} G'(z_\textrm{s})}{G(z_\textrm{s})},
\label{eq_saddle}
\end{align}
where
\begin{align}
G(z_\textrm{s}) 
&= 1 + z_\textrm{s} + \sum_{n=2}^\infty \left[ \frac{1}{\ln n!} \left( 1 - \frac{1}{n!} \right)\right] z_\textrm{s}^n
\label{eq_g1}
\end{align}
and
\begin{align}
G'(z_\textrm{s}) 
&= 1 + \sum_{n=2}^\infty \left[ \frac{1}{\ln n!} \left( 1 - \frac{1}{n!} \right)\right] n  z_\textrm{s}^{n-1}.
\label{eq_g2}
\end{align}
Equations~\eqref{eq_g1} and \eqref{eq_g2} are easily obtained from eq.~\eqref{eq_g0}.
From the above calculations, we obtain the following expression by rearranging eq.~\eqref{eq_ln_f}:
\begin{align}
&\left\langle \ln f_{k}^{(\beta_1)} \right\rangle_{\{ \beta_2, \dots, \beta_N\}} \nonumber \\
&\quad = (\beta_1 - 1) \ln k! - \ln H(z_\textrm{s},\beta_1) + k \ln z_\textrm{s}.
\label{eq_av_ln_fk}
\end{align}
Therefore, the equilibrium occupation distribution of urn $1$ with $\beta_1$ 
in the configurational average is obtained 
from eqs.~\eqref{eq_p_k_0}, \eqref{eq_p_k}, and \eqref{eq_av_ln_fk} as follows:
\begin{align}
P(k, \beta_1)
&= (k!)^{\beta_1-1} \frac{z_\textrm{s}^k}{H(z_\textrm{s},\beta_1)} \nonumber \\
&= \frac{(k!)^{\beta_1-1} z_\textrm{s}^k}{\sum_{n=0}^\infty (n!)^{\beta_1-1} z_\textrm{s}^n}.
\end{align}
Finally, we obtain the equilibrium occupation distribution of the whole system as 
\begin{align}
P(k) 
&= \int_0^1 \textrm{d}\beta \,\, \frac{(k!)^{\beta-1} z_\textrm{s}^k}{\sum_{n=0}^\infty (n!)^{\beta-1} z_\textrm{s}^n}.
\label{eq_distribution}
\end{align}
Note that the equilibrium occupation distribution is suitably normalized.

A large density $\rho$ leads to $z_\textrm{s} \simeq 1$,
and hence we consider that the saddle-point is approximately $z_\textrm{s} = 1$ for $\rho \gg 1$.
Substituting $z_\textrm{s} = 1$ into eq.~\eqref{eq_distribution}, 
we obtain the asymptotic form of the equilibrium occupation distribution as follows (see Appendix~A):
\begin{align}
P(k) &\sim k^{-2} \frac{1}{(\ln k)^2}.
\label{eq_asymptotic}
\end{align}
Hence, the equilibrium occupation distribution follows a generalized power law
with a squared inverse logarithmic correction.

\begin{figure}
\begin{center}
  \includegraphics[width=8cm,keepaspectratio,clip]{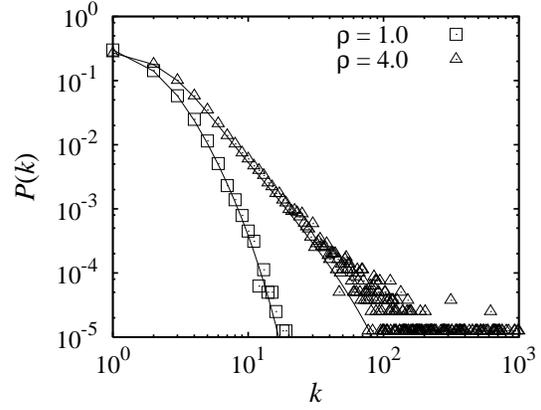} 
\caption{
Equilibrium occupation distributions of the preferential urn model.
The numerical results for the density $\rho = 1.0$ are shown by squares,
and those for $\rho = 4.0$ by triangles.
The solid lines correspond to eq.~\eqref{eq_distribution} for respective cases.
In both cases, the number of urns $N$ is $1000$, and that of the transfer process of balls is $2000N$.
The data are averaged over $20$ different realizations.
}
\label{fig_results}
\end{center}
\end{figure}

To confirm the above analytical treatments, we performed numerical experiments.
Because initial distributions are arbitrary, we put each ball to urns at uniformly random as initial configurations.
In the numerical experiments,
the number of urns $N$ is $1000$,
and that of the transfer process of balls is $2000N$.
We checked that the $2000N$ transfer processes are enough to give the equilibrium states.
Figure~\ref{fig_results} shows the comparison between the equilibrium occupation distributions
of numerical experiments and those of the replica analysis.
For the density $\rho = 1.0$, we obtain the saddle-point $z_s = 0.660$ from eq.~\eqref{eq_saddle};
$z_s = 0.967$ for $\rho = 4.0$.
The numerical experiments indicate that the replica analysis is adequate.

\section{Discussions}
\subsection{Intuitive understanding of the local temperatures}
Here, we discuss an intuitive understanding of the preferential urn model.
The heterogeneity of the inverse local temperatures indicates
that each urn has different ability to obtain balls.
The inverse local temperature allows us to obtain the following intuitive picture;
urns with high-temperature tend to kick out balls,
and those with low-temperature are likely to stock balls.
Therefore, when there is no randomness, 
all urns compete for balls, so that a fat-tailed behavior does not occur.
On the contrary, when an urn has a lower-temperature,
the urn tends to get a lot of balls.
Hence, it is easy to imagine that urns with a lot of balls can emerge due to randomness.

\subsection{Restriction for the inverse local temperature $\beta$}

We note that the value of the inverse local temperature $\beta_i$ has a restriction:
the value of $\beta_i$ can not be over $1$.
For example, eq.~\eqref{eq_g0} does not converge if $\beta > 1$.
The restriction of $\beta < 1$ means that the urn with $\beta > 1$ is too cold to kick out balls in the urn,
so that the model does not work well.
Actually, we have confirmed by numerical experiments that almost all balls concentrate in an urn with $\beta > 1$.
Therefore, the restriction of $0 \leq \beta \leq 1$ is necessary.

\subsection{Annealing systems}

As mentioned in \S~3,
it is difficult to calculate the configurational average of the numerator and denominator
at the same time, i.e., the configurational average of eq.~\eqref{eq_p_k_0}.
Hence, we use the replica method to calculate the configurational average.
The difficulty stems from the fact that the preferential urn model has \textit{quenched disorder};
evaluation of a physical quantity is taken for a given fixed (quenched) set of $\beta_i$,
so that the inverse local temperatures are time-independent.
However, if we consider an annealing system, i.e., the inverse local temperatures are not quenched
and changed with time enough fastly,
we can easily take the average of eq.~\eqref{eq_p_k_0},
and there is no need to use the replica method.

For the annealing system, 
\begin{align}
\left\langle \frac{Z_2}{Z_1}  \right\rangle_{\{ \beta_2, \dots, \beta_N \}}
= \frac{\left\langle Z_2 \right\rangle_{\{ \beta_2, \dots, \beta_N \}}}
{\left\langle Z_1  \right\rangle_{\{ \beta_2, \dots, \beta_N \}}},
\end{align}
and these quantities of $\left\langle Z_1 \right\rangle_{\{ \beta_2, \dots, \beta_N \}}$
and $\left\langle Z_2 \right\rangle_{\{ \beta_2, \dots, \beta_N \}}$
are easily obtained.
Finally, we can get the equilibrium occupation distribution for the annealing system
as the same form of eq.~\eqref{eq_distribution}.
It is worth pointing out that
the case of the annealing system gives the same results as for the quenched system.
For the annealing system,
we can straightforwardly get the equilibrium occupation distribution without using the replica method,
then one might consider that there is no need to use replica method to analyze urn models with randomness.
However, when we have performed numerical experiments for the annealing system
in which all inverse local temperature vary at each time step,
the fat-tailed behavior does not occur.
Hence, it is an open question why the result for the annealing system agree in that for the quenched system,
and how situation corresponds to the annealing system.
At least, we consider that the replica method is adequate for analyzing urn models with quenched randomness.

\subsection{Econophysical application}

The preferential urn model stems from the analysis of the nongrowing models for complex networks \cite{Ohkubo2005c},
but it would be a useful model for various phenomena.
For example, we consider an econophysical model in which many agents exchange their wealth.
Each agent corresponds to each urn,
and the transfer procedure of their wealth is represented by the transfer process of balls.
We here assume the transition rate $W_{k_i \to k_i+1} \propto (k_i+1)^{\beta_i}$,
where $k_i$ is the amount of wealth of agent $i$, 
and $\beta_i$ represents an ability of agent $i$ for obtaining wealth.
The above dynamics indicates the following statements:
\begin{itemize}
\item A wealth to be transferred is selected at random. 
It means that agents with large wealth are likely to loose their wealth
(Agents with large wealth have a lot of opportunity to consume their wealth.)
\item Agents with larger wealth have the chance to get wealth in higher probability.
\item Each agent has a different ability for obtaining wealth.
\end{itemize}
Hence, the above model seems reasonable for an econophysical one.
The econophysical model can be treated analytically, as shown in the present paper;
supposing that the distribution of their abilities $\beta_i$ is an uniform one,
we conclude that the wealth distribution has a fat-tailed form of eq.~\eqref{eq_asymptotic}
when total wealth in the whole system is enough large.
Though the fat-tailed form has the squared inverse logarithmic correction,
the cumulative distribution of eq.~\eqref{eq_asymptotic} is similar to the Parato's distribution\cite{Newman2005}.
The preferential urn model has not been proposed in order to the Parato's distribution,
so that it is unclear that the above discussion is suitable; the model is just a trial one.
However, it seems reasonable to assume that a system includes some randomness;
each agent is in various environments and has a different ability.
Furthermore, the preferential concept also seems to be adequate for real econophysical systems.

\section{Concluding Remarks}

In summary, we analyzed the preferential urn model with randomness using the replica method.
While it is in general difficult to analyze urn models with randomness,
the preferential urn models with uniform randomness for inverse local temperatures
can be treated analytically by using the replica method.
From the analysis,
it was revealed that the equilibrium occupation distribution has a fat-tailed form when the density is enough large.
The fat-tailed distribution does not follow a pure power-law distribution;
the distribution has a generalized power law with a squared inverse logarithmic correction.

The preferential urn model would produce various further studies,
such as nonequilibrium properties, other variation of the urn models with randomness,
and applications of the urn models for various other research fields.
It will also be important to discuss relationships with interacting particle systems,
such as the zero-range process \cite{Evans2005}.

\section*{Acknowledgment}
The authors thank Naoki Masuda for his helpful comments on the manuscript.
This work was supported in part by grant-in-aid for scientific research (No. 14084203 and No. 17500134)
from the Ministry of Education, Culture, Sports, Science and Technology, Japan.
J. O. was partially supported by Research Fellowship from Japan Society
for the Promotion of Science (JSPS) for Young Scientists.

\appendix
\section{Derivation of eq.~\eqref{eq_asymptotic}}
When we approximate the saddle-point $z_\textrm{s} = 1$ for $\rho \gg 1$,
the equilibrium occupation distribution of eq.~\eqref{eq_distribution} is given by
\begin{align}
P(k) &= \int_0^1 \textrm{d}\beta\frac{(k!)^{\beta-1}}{\sum_{n=0}^\infty (n!)^{\beta-1}}. 
\end{align}
By using the partial integration twice,
the approximate equilibrium occupation distribution for large $k$ region is obtained as follows:
\begin{align}
P(k) 
&= \frac{1}{\ln (k!)} \left[ \frac{1}{\sum_{n=0}^\infty 1} - 
  e^{-\ln k!}\frac{1}{\sum_{n=0}^\infty (n!)^{-1}} \right] \nonumber \\ 
& \quad + \frac{1}{(\ln (k!))^2} \left[ \frac{H'(1,1)}{H(1,1)^2} 
-  e^{- \ln (k!) } \frac{H'(1,0)}{H(1,0)^2} \right] \nonumber \\
& \quad + \frac{1}{O( (\ln (k!))^3  )} \nonumber  \\
&\sim \frac{1}{(\ln (k!))^2} \frac{\sum_{n=0}^\infty \ln (n!)}{\left( \sum_{n=0}^\infty 1 \right)^2} \nonumber \\
& = \frac{1}{(\ln (k!))^2} 
\lim_{S\to \infty }\frac{\sum_{n=0}^S \ln (n!)}{\left( \sum_{n=0}^S 1 \right)^2} \nonumber \\
& \sim \frac{1}{(\ln (k!))^2} \lim_{S\to \infty }\frac{O(S^2)}{O(S^2)} \nonumber \\
& \sim k^{-2} \frac{1}{(\ln k)^2},
\end{align}
where we use the following relation \cite{Prudnikov1986},
\begin{align}
\lim_{S \to \infty} \sum_{n=0}^S \ln (n!) \sim \lim_{S \to \infty } O(S^2),
\end{align}
and the Stirling formula,  $\ln (k!) \sim k \ln k - k$.


\end{document}